# Bose-Einstein Condensation of Gases in the Frame of Quantum Electrodynamics: Interconnection of Constituents

Mark E. Perel'man [1)]
*Racah Institute of Physics, Hebrew University, Jerusalem, Israel*

Bose-Einstein condensate of rarified atomic gases is considered as the state formed by exchange of virtual photons, resonant to the lowest levels of atoms; such representation corresponds to the Einstein opinion about an inter-influence of condensable particles. Considered interactions directly lead to the QED structure of nonlinear potential in the Gross-Pitaevskii equation. Linear momenta that correspond to the thermal energy of condensable atoms are connected to near field of particles and therefore leave atoms immovable. The estimations of these effects do not contradict the observed data; the general quantum principles predict possibility of stimulating of BEC formation by resonant irradiation. All this requires the spectroscopic investigation of BEC on different steps of formation.

## 1. Introduction

The realization of Bose-Einstein condensation (BEC) in rarefied atomic gases ([[1], [2]], references on concrete gases is given below, in the Table) for the first time had given possibilities for detailing the kinetics of the phenomenon: really, in the earlier condensates (quantum liquids) the condensable and non-condensable particles are so mixed that a detachment of BEC particles interaction is difficult, if in general possible.

The big density of particles in the volume of condensates had led, naturally, to the description of their properties through pair interactions, via amplitudes of elastic scattering. Theoretically the processes of BEC of gases are intensively studied with the statistical point of view, by the famous Gross-Pitaevskii equation (GPE, the comprehensive reviews [[3], [4]]).

Note, that although the concept of possible condensation of quantum particles in the momentum space was entered by Einstein completely statistically [[5]], in the second part of these articles he had already pointed out that in such state should appear the inter-influence of molecules, this Einstein opinion is supported by some investigators [[6], [1]].

To these common remarks must be added that as the density of atomic gases is so small that interatomic collisions can be neglected, it is consequently necessary to consider interaction (interconnection) of condensable particles only and only through their near fields, a field of (virtual) photons, i.e. in the frame of QED.

Besides absence of the account of interconnection of condensable particles, the cited theory is not free from certain physical contradictions. So, if BEC consist from atoms with non-zero temperature, these atoms should have definite linear momenta; however it is postulated that all momenta of atoms are strongly equal to zero. But the general relativistic ansatz requires the appearance of momentum if energy exceeds inertial mass.

This contradiction is naturally resolved in the frame of QED: physical object is the atom connected with the own near field, therefore their energy and momentum can not be separated (formally the renormalization of mass can be considered). BEC means the association of near fields of atoms, within which the atom absorbs virtual photons (isotropic in a first approximation), promptly reemits them into different directions and therefore remains at rest on the average. The joining of near fields up allows the consideration of condensate as a uniform object and consequently its description by one $\Psi$-function.

---
[1]). E-mail: m.e.perelman@gmail.com.

Such approach requires the reexamination of BEC kinetics in the frame of QED. Notice that the GPE, which satisfactory describes many features of BEC, includes the nonlinear potential that must be qualified in the scope of QED and in accordance with the experimental data.

What's the news can give the QED approach to the field besides précising of interconnections between particles? So, in particular, the quantum theory of radiation can predict possibility of stimulated formation of BEC by resonant irradiation; can help in examination of BEC formations by different species in dependence on conformity of their spectra, etc.

As a heuristic reason to the microscopic description such supervision could be taken into account: the potential of the GPE is proportional to the density of particles, $|\Psi|^2 \sim r^{-3}$, but such kind has, as distinct from the van-der-Waals potential, the resonant interaction between identical neutral atoms. Already this circumstance can be considered as a prediction of an opportunity of resonant interaction between particles in BEC.

Notice that phase transitions within the QED approach to theory of condensed substances can be described via changing of durations of virtual photons exchange that execute the bonds of substance constituents [7]; therefore the removal of latent heat of the first kind transitions leads to emission of corresponding frequencies [8] (general theory of durations of elementary acts is given in [9]). Constituents of substance in the QED approaching must be considered jointly with their near field [10], it means that internal changing at transitions can be attributed, partially or even completely, to a changing of near field.

In our cited examinations only the energy of bonds had been considered, but it is not excluded that virtual quanta of near field can transfer momenta in tunnel regime. In other words it is not excluded that momenta of BEC atoms can be uniformly distributed in the near field and therefore atoms can remain immovable. Such hypothesis requires the executing of strict resonance conditions, i.e. the existence of proper (dipole) transitions and the strict correspondence of thermal wave lengths and distances between atoms. In the Section 2 will be shown that all such necessary correspondences can be real and can allow the comparison with experimental data.

In the Section 3 the deducing of potential of resonant exchanging is executed. It just gives the term, addition of which to the Schrödinger equation turns it into the GPE. Results and certain perspectives are summed in the Conclusions.

**2. BEC formation**

Let us consider the BEC formation as the result of existence of exchange forces that bind the distributed systems of oscillators. In the frame of QED the exchange corresponds to emission of resonant frequencies carrying away energy and/or kinetic momenta of particles and their virtual reabsorption by other scatterers. Kinetics of such processes is described via S-matrix of scattering by two temporal functions, the time delay at scattering and the time duration of free photon formation [7, 9]:

$$\tau \equiv \tau_1 + \tau_2 = \frac{d}{id\omega} \ln S . \qquad (1)$$

For two-level system with upper level ($\omega_0$, $\Gamma$) and excitation energy $\hbar\omega \sim \kappa T$, such that $\omega \ll \omega_0$, they are expressed, correspondingly, as

$$\tau_1 = \frac{\Gamma/2}{(\omega-\omega_0)^2 + \Gamma^2/4} \to \frac{\Gamma}{2\omega_0^2}; \qquad \tau_2 = \frac{(\omega-\omega_0)}{(\omega-\omega_0)^2 + \Gamma^2/4} \to \frac{1}{\omega_0}. \qquad (2)$$

It leads to the main restriction on density of the considered medium n ~ $\lambda^{-3}$, which should be close to the condition of Bose condensate formation defined by Einstein [5] as

$$n\Lambda^3 = 2.615, \qquad (2)$$

n is the density of identical particles (bosons), and $\Lambda = \hbar\,(2\pi/m\,\kappa T)^{1/2}$ is the length of de Broglie thermal wave.

The combination of these conditions leads to our *main assumption*: BEC of rarified atomic gases can be formed under coordination of thermal excitation (thermal wave length) and the energy of sufficiently effective lowest level of atom (the ns→np transition): $\Lambda \sim \lambda$.

It means that particles virtually passing into the condensate at this "phase" transition will be characterized by zero momenta; momenta of virtual quanta of near field at absorption and reemission will be of the order of $p = 2\pi\hbar/\Lambda$ with $pc \gg \kappa T$, i.e. processes of quanta transferring are of tunneling type and can be instantaneous (cf. [11]). Thus the existence of a certain field transmitting interaction between particles in the condensate, but distinct from the van-der-Waals forces, is implicitly accepted.

Let's consider this assumption in more details. Each atom has, on the average, the kinetic energy $E_1 = p_1^2/2m = \xi\kappa T$ ($\xi$ is the coefficient of proportionality; probably, in view of condensation [5], $\xi = 1.348/2.615 \sim 1/2$, its magnitude introduce certain uncertainty, but it is not very important for further examination). Therefore at entering into the condensate each atom must get rid of corresponding momentum,

$$p_1 = (2mE_1)^{1/2} \to (2\xi m\,\kappa T)^{1/2}, \qquad (3)$$

by atomic collisions, which are very rare at the considered situation, or by a transfer of this momentum via a virtual photon emission to any other constituent.

The energy of photon with the momentum (3) may correspond therefore to the wavelength of the strongest level of recipient atom $\lambda_0$, i.e. $\lambda_0$ should be of the order of

$$\lambda_0 = 2\pi\,\hbar/p_1 = 2\pi\,\hbar/(2\xi m\,\kappa T)^{1/2}. \qquad (4)$$

Additionally we must accept that the transfer of this excitation would be the most effective on distance of the photon wavelength, i.e. if all interacting atoms of BEC are in the near field of exchanging photons:

$$\lambda_0 \sim n^{-1/3}. \qquad (2')$$

. In the Table are extracted the most suitable, as seems, lines, with which are calculated via (4) and (2') the temperatures and densities, then they are compared with the experimental data.

The comparison shows a qualitative accordance of the data for Na, Cr and Rb or, more correctly, an absence of essential contradictions.

Some evident discrepancies can be mentioned and discussed. So, for H are possible two mechanisms: the Lyman line is insufficiently strong and therefore the BEC could consist from two parts. In the case of Li condensable atoms must overcome the repulsion connected with negative sign of scattering amplitude and it must naturally lower the temperature of transition. In the case of Cs the third particle for transition executing is needed (the Efimov mechanism), etc.

THE TABLE

| species | $\lambda_0$ μm | $T_{calc}$ μK | $n_{calc}$ cm$^{-3}$ | $T_{exp}$ μK | $n_{exp}$ cm$^{-3}$ | References |
|---|---|---|---|---|---|---|
| $^1$H | 0.121 (L$_\alpha$) | 1260 | 5.64×10$^{14}$ | | 1.8×10$^{14}$ | [12] |
| | 0.656 | 429 | 3.5×10$^{12}$ | 50 | | |
| $^7$Li | 0.671 | 5.86 | 3.31×10$^{12}$ | 0.30 | 1.5×10$^{12}$ | [13] |
| $^{23}$Na | 0.590 | 2.32 | 4.87×10$^{12}$ | 2.0 | 1.5×10$^{14}$ | [14] |
| $^{52}$Cr | 0.425 | 1.96 | 1.3×10$^{13}$ | 0.7 | 1.2×10$^{12}$ | [15] |
| $^{87}$Rb | 0.780 | 0.35 | 2.11×10$^{12}$ | 0.17 | 2.5×10$^{12}$ | [16] |
| | 0.795 | 0.336 | | 0.67 | 2.2×10$^{14}$ | [17] |
| $^{133}$Cs | 0.851 | 0.19 | 1.62×10$^{12}$ | 0.01÷0.2 | 10$^{11}$÷10$^{13}$ | [18] |

The offered approach can be further generalized. So the new problems and new possibilities for examinations have been appeared with the mixtures cooling (e.g. [19]): the dynamics of sympathetic cooling, the interspecies scattering properties, possibilities of resonant transferring of excitations and so on must be investigated. But at all cases the examination of offered mechanism strongly requires the spectroscopic verifications.

Consecutive absorptions and reemissions of resonant quanta lead to the induced radiation [20]. If such acts are essential for BEC existence, in this formation should be noticeable stimulated processes of radiation. Therefore it is possible to think that the resonant irradiation of advanced samples can speed up the BEC formation.

### 3. Interaction of atoms in near field and nonlinear potential

Energy of nonresonant interaction of two neutral atoms in the near field is determined by the two-photon exchange (the fourth order of S-matrix) as [21]:

$$U(\mathbf{r}) = (i/4\pi) \int_{-\infty}^{\infty} d\omega\, \omega^4\, \alpha_1(\omega)\alpha_2(\omega)[D_{ik}(\omega, \mathbf{r})]^2, \qquad (5)$$

where $\alpha_i(\omega)$ is the polarizability of cooperating atoms, scalar for atoms in the S-state.

The Green functions of the wave equation can be decomposed (e.g. [22]) as

$$D_{ij}(\omega, \mathbf{r}) = \{(\delta_{ij} + e_i e_j) - (i/\omega r)\, P_{ij}\, \text{ctg}(\omega r) + (1/\omega r)^2\, P_{ij}\} D(\omega, \mathbf{r}) \qquad (6)$$

with directing cosinuses $e_i = x_i/r$ and the tensor $P_{ij} = \delta_{ij} - 3 e_i e_j$.

Three terms of (6) appropriate to far, intermediate (transient) and near fields, are represented, accordingly, in the (t, $\mathbf{r}$)-picture via the Pauli-Jordan function:

$$D_{ij}(t, \mathbf{r})|_{FF} = (\delta_{ij} + e_i e_j)\, D(t, \mathbf{r});$$

$$D_{ij}(t, \mathbf{r})|_{MF} = (1/4\pi r)\, P_{ij}\, \theta(r^2 - t^2) \equiv r^{-1} P_{ij}\, \partial_t D_N(t, \mathbf{r});$$

$$D_{ij}(t, \mathbf{r})|_{NF} = (1/4\pi r^2)\, P_{ij}\, \{\text{sgn}(t)\theta(t^2 - r^2) + (t/r)\theta(r^2 - t^2)\} \equiv r^{-2} P_{ij} D_N(t, \mathbf{r}), \qquad (7)$$

where $\theta(t)$ is the Heaviside step function, and

$$D_N(\omega, \mathbf{r}) = -\omega^{-2} D(\omega, \mathbf{r}) = -(1/2\pi i \omega^2 r) \sin(\omega r) \qquad (8)$$

corresponds to the function used by Schwinger in [23].

If to substitute the decomposition (6) into (5) (note the affinity of $D_{il}(\omega, \mathbf{r})|_{NF}$ to matrix element of dipole-dipole interaction) with the precise repeating of the procedure [21], it results in the van-der-Waals energy of interaction proportional $R^{-6}$, and the energy of Casimir interaction of atoms proportional to $R^{-7}$. Thus they are described by the propagator (8), i.e. they occur in the near field and, at least in part, can be transferred superluminally.

However for resonant interaction between identical (motionless) atoms,

$$A^*_1 A_2 \leftrightarrow A_1 A^*_2, \qquad (9)$$

matrix element is nonzero still in the second order:

$$S^{(3)} = -\tfrac{1}{2} \int dt_1 dt_2 \, T\{V(t_1)V(t_2)\}, \qquad (10)$$

where $V = -\mathbf{E}(\mathbf{r}_1)\mathbf{d}_1 - \mathbf{E}(\mathbf{r}_2)\mathbf{d}_2$. Therefore instead of (5) we have

$$U(\mathbf{r}) = (i/2\pi) \int_{-\infty}^{\infty} d\omega \, \omega^2 \, D_{ik}(\omega, \mathbf{r}) \, \mathrm{Re}[\alpha_{ik}(\omega)], \qquad (11)$$

with the tensor of scattering of two-level, for simplicity, systems, expressed through matrix elements of dipole moments:

$$\alpha_{ik}(\omega) = (d_i)_{01} (d_k)_{10} /(\omega_0 - \omega - i\Gamma) + (d_k)_{01} (d_i)_{10} /(\omega_0 + \omega - i\Gamma).$$

At substitution the Green functions of near field in (11) it is received that interaction decreases in this case as $R^{-3}$ (cf. [24]) and can be expressed via the amplitude of s-scattering. On the other hand it means that the potential of interaction (11) is proportional to $|\psi(x)|^2$ and just represent the nonlinear term of the GPE, i.e. of the "cubical" Schrödinger equation.

The full probability of process (9) is determined in the near field as

$$W \sim \int_{-\infty}^{\infty} d\omega \, |D_{ik}(\omega, \mathbf{r})\alpha_{ik}(\omega)|^2 \to \int_{-\infty}^{\infty} d\omega \, |\mathbf{d}_1|^2 \, |\mathbf{d}_2|^2 |D_{ik}(\omega, \mathbf{r})|_{NF}|^2 \tau_1(\omega)/\Gamma. \qquad (12)$$

Carrying out integration in view of δ-character of $\tau_1$, using matrix elements of dipole operators $|\mathbf{d}|^2 = \hbar e^2 f/2m\omega$, where *f* is the oscillator force, and substituting the expressions of singular functions (8), we receive that the probability of process depends on distance between cooperating atoms as $R^{-6}$, i.e. takes the form of well-known half-empirical Förster law [25] (see, e.g., [26]):

$$W = \Gamma^{-1} (R_0/R)^6, \qquad (14)$$

where $R_0$ is so-called Förster radius.

From (12) follows that the rate of process (9) in the time representation is represented by the square of near field singular function (8):

$$|D_{il}(t, \mathbf{r})|_{NF}|^2 = (1/4\pi r^2)^2 \{\theta(t^2 - r^2) + (t/r)^2 \theta(r^2 - t^2)\}, \qquad (15)$$

that determines relative probabilities of excitation transfer with subluminal and superluminal speeds.

## Conclusions

Our considerations can be summed as follows.

The Bose-Einstein condensate of rarified atomic gases BEC can be described as formation of such set of atoms, kinetic energy and momenta of which are concentrated in their coexisted near field, its virtual. quanta should be continuously absorbing and promptly reemitting by atoms. The isotropy of such acts effectively leads to motionlessness of atoms and therefore these properties can be taken as the definition of BEC.

Two approaches to the description of BEC of rarified atomic gases, the statistical and in the frame of quantum electrodynamics, are supplemental to each other. In particular, QED determines a kind of the potential necessary for the generalization of the Schrödinger equation up to the Gross-Pitaevskii equation.

Checking of the considered phenomena would be carried out by examination of BEC radiation field. Essentially interesting seems the opportunity of stimulated BEC formation by an external resonant irradiation.

The offered approach can and should be generalized on cases of gases mixtures, and also on Fermi-gases. The possibilities of analogical approach to BECs of another types is not discussed, but a needness of searching of possibilities of particles interconnections would become more evident.